\documentclass[aps,pra,amsmath,amssymb,twocolumn,superscriptaddress,showpacs]{revtex4}
\usepackage{amsmath,graphicx,float,epsfig,float,tabularx,multirow,times}
\usepackage{times,epstopdf}
 
\newcommand{\Tr}[2]{\mathrm{Tr}_{#1}\left[ {#2} \right]} 
\newcommand{\ket}[1]{\left\vert {#1} \right\rangle} 
\newcommand{\bra}[1]{\left\langle {#1} \right\vert}

\newcommand{\csch}{\mathrm{csch}}
\newcommand{\smi}{{\scriptscriptstyle I}}
\newcommand{\nbar}{{\overline n}}
\begin{document}
\title{Qubit-assisted thermometry of a quantum harmonic oscillator}
\author{Matteo Brunelli}
\affiliation{Dipartimento di Fisica, Universit\`a degli Studi di Milano, I-20133 Milano, Italy}
\author{Stefano Olivares}
\affiliation{Dipartimento di Fisica, Universit\`a degli Studi di Milano, I-20133 Milano, Italy}
\affiliation{CNISM, UdR Milano, I-20133 Milano, Italy}
\author{Mauro Paternostro}
\affiliation{Centre for Theoretical Atomic, Molecular and Optical Physics, 
School of Mathematics and Physics, Queen's University Belfast, 
Belfast BT7\,1NN, United Kingdom}
\author{Matteo G. A. Paris}
\affiliation{Dipartimento di Fisica, Universit\`a degli Studi di Milano, I-20133 Milano, Italy}
\affiliation{CNISM, UdR Milano, I-20133 Milano, Italy}
\begin{abstract}
We use the theory of quantum estimation in two different qubit-boson
coupling models to demonstrate that the temperature of a quantum
harmonic oscillator can be estimated with high precision by
quantum-limited measurements on the qubit. The two models 
that we address embody situations of current physical interest due to
their connection with ongoing experimental efforts on the control of
mesoscopic dynamics. We show that population measurements performed over
the qubit probe are near optimal for a broad range of temperatures of
the harmonic oscillator.
\end{abstract}
\date{\today}
\maketitle
\section{Introduction}
The improved control over systems of intrinsic complexity makes the
implementation of techniques for the inference of specific properties of
their states a necessary step towards the achievement of full quantum
control. Yet, it is often the case that the device into which we would
like to enforce quantum mechanical features is not fully or easily
addressable. System-interrogation can only be performed, in such cases,
in an indirect way through the use of probes of an appropriate
nature~\cite{campbell2010}.  It is thus very important to devise
experimentally implementable strategies for the inference of properties
of inaccessible quantum systems, identify the optimal state-preparation
of the probe as well as the observable that allows for the maximum
extraction of information about the parameter that we are interested in.
\par This agenda is dressed of even more relevance due to the recent
experimental efforts produced towards the quantum-limited management of
mesoscopic systems, such as superconducting devices~\cite{dicarlo},
light-interfaced cold-atom systems~\cite{hammerer} and mechanical
systems operating at the quantum level~\cite{aspelmeyer}. All such
systems have in common the use of  ``quantum interfaces" with devices of
a different nature, which are then exploited for state-preparation,
manipulation of information and possibly read-out. The quantum-interface
paradigm is indeed very fruitful for the extraction of information out
of a system that is only partially accessible: through the coupling with
a controllable subsystem, one can indeed arrange for mechanisms able to
provide useful knowledge on key features of a dynamics or a state.
Examples of such a possibility, which have been materialised in
successful experimental demonstrations, include the micro-maser
technology for the revelation of the properties of the field within a
high-quality microwave cavity~\cite{haroche}, the coupling of a
Bose-Einstein condensate to a (classical) mechanical oscillator for the
investigation on Casimir-Polder effects~\cite{hunger}, and intra-cavity
quantum optomechanics, where the radiation-pressure force is used to
read the noise properties of a mechanical mode~\cite{mauroNJP}. In the
latter context, in particular, a key parameter is embodied by the
temperature at which the mechanical mode operates. Indeed, unwanted
thermal effects typically spoil the quantum features enforces in the
mechanical system by means of a coherent evolution. Having a precise
quantitative estimate of the entity of such effects \cite{Pir11} would be crucial not
only for prediction purposes but also to design in the best possible way
a quantum-enforcing protocol that accounts, ab initio, such undesired
effects. Needless to say, these considerations can be extended
straightforwardly to any of the scenarios addressed above.  
\par 
Recently, strategies for the determination of the temperature of a
harmonic oscillator have been put forward, based on the coupling to a
quantum probe embodied by a two-level system (a
qubit)~\cite{Bru11,lovett2012}. The coupling model to be sued for the
thermometry of the oscillator's state was the Jaynes-Cummings one,
within and beyond the so-called rotating wave approximation~\cite{JC}.
While Ref.~\cite{lovett2012} proposed the use of the ac Stark effect as
a way to infer the temperature of the oscillator, Brunelli {\it et
al.}~\cite{Bru11} have applied the proper tools of quantum estimation
theory (QET)~\cite{lqe1,lqe2,lqe3} to design optimal protocols for the estimate
of the system's temperature.  
\par 
In this paper, we significantly extend the approach in~\cite{Bru11} to
other physically motivated qubit-oscillator models, proving that optimal
and effective thermometry can indeed be performed by means of simple
measurements onto the qubit's state. We tackle both the
coupling between a superconducting qubit and a nano electromechanical
oscillator and the far-off resonant interaction between a two-level atom
and the field of a cavity, thus providing an analytical QET-based study
of an ample spectrum of experimentally motivated situations.  
\par 
The remainder of this work is
organized as follows: in Sec.~\ref{modello} we describe the general
system that we address and introduce the QET tools for our analysis.
Sec.~\ref{modello1} studies the first model of our investigation, which
addresses the capacitive coupling of a superconducting qubit and a
nano-electromechanical oscillator. In Sec.~\ref{modello2} we assess our
QET-based approach in the case of a qubit that is off-resonantly coupled
to a harmonic oscillator, such as for a two-level atom in a far-off
resonant cavity. In both instances, population measurements over the
probing qubit allow for the optimized estimate of the oscillator's
temperature. Finally, in Sec.~\ref{conc} we summarize our findings and
open up new perspectives.
\section{Approach to the problem}
\label{modello}
Let us consider a general quantum harmonic oscillator with frequency
$\Omega$ and  at thermal equilibrium with its environment. The state of
the oscillator is described by the Gibbs density operator (we use
natural units, i.e. $\hbar=1$, throughout the manuscript) 
\begin{equation}
\rho_o=\frac{e^{-\beta\Omega\, a^\dag a} }{{\cal Z}} =
\sum_{n=0}^{\infty}\frac{\nbar^n}{(\nbar +
1)^{n+1}}\ket{n}\bra{n},    
\end{equation}
where $\nbar=(e^{\beta \Omega}-1)^{-1}$ is the average number of thermal
excitations, $\ket{n}$ is a Fock state with $n$ quanta, ${\cal
Z}=\mathrm{Tr}[e^{-\beta\Omega\, a^\dag a}]$ is the partition function
and $\hat a$ ($\hat a^\dag$) is the bosonic annihilation (creation)
operator of the harmonic oscillator.  Our aim is to estimate the inverse
temperature $\beta=1/k_BT$ of the oscillator by coupling it with a qubit
encoded in the logical states $\{\ket{0}_q,\ket{1}_q\}$ of a two-level
system that is initially prepared in the general pure state
\begin{equation}
\ket{\psi}_q=\cos \frac{\theta}{2} \ket{0}+
e^{i\varphi}\sin \frac{\theta}{2}\ket{1}.
\end{equation}
Here $(\theta,\varphi)$ are the angles defining the orientation of the
qubit's Bloch vector in the corresponding Bloch sphere, while $k_B$ is
the Boltzmann constant.
We assume no initial correlation between the
probe and the oscillator and also assume that the interaction Hamiltonian 
has the general form 
\begin{equation}
\hat H_{\smi}=g \, {\hat A}_o\otimes\hat A_q\,,
\end{equation} 
where ${\hat A}_o$ ($\hat A_q$) is an operator in the Hilbert space of
the oscillator (qubit) and $g$ a coupling constant. In what follows, we
shall call $\{\ket{x}_o\}$ a basis of states of the harmonic oscillators
that are eigenstates of $\hat A_o$, i.e. $\hat A_o\ket{x}_o=x\ket{x}_o$.
\par
Any measurement aimed at estimating the temperature of the oscillator is
performed on the {state} $\varrho_q$ of the probing qubit after its
joint evolution with system $o$. That is 
\begin{equation} 
\begin{aligned}
\varrho_{q}(\beta)&= 
\Tr{o}{\hat U\,|\psi\rangle_q{}_q\langle\psi| 
\otimes\rho_o\, \hat U^{\dagger}} \notag 
\\ &=\int\!\! dx\, \rho_o(x) 
e^{-igt x\hat A_q}
|\psi\rangle_q{}_q\langle\psi|
\, e^{igt x \hat A_q}
\end{aligned}
\end{equation}
where $\rho_o(x)={}_o\langle{x}|\rho_o|x\rangle_o$ are the diagonal
matrix elements of the initial thermal state in the basis $|x\rangle_o$
of the oscillator operator $\hat A_o$. In what follows, we make use of
the apparatus of QET to design the optimal probing state and
measurements needed to estimate the inverse temperature $\beta$.
According to the Cram\'er-Rao inequality, the variance $\delta(\gamma$)
of any unbiased estimator of an arbitrary quantity $\gamma$ satisfies
the inequality 
\begin{equation}
\label{bound}
\delta(\beta)\ge\frac{1}{MF(\beta)}
\end{equation}
with $M$ the number of measurements used in order to perform the
estimate and $F(\beta)$ the Fisher information of $\beta$, which is
defined as 
\begin{equation}
F(\beta)=\sum_{j}p_{j}(\partial_\beta\ln p_{j})^2=
\sum_{j}\frac{|\partial_\beta p_j|^2}{p_j},
\end{equation}
where $p_{j}$ is the probabilities to get outcome $j$ from a
measurements performed over the qubit state and described, in general,
by the positive operator valued measurement (POVM) $\{\hat\Pi_j : \hat
\Pi_j\ge0, \sum_j\hat \Pi_i=\openone\}$. Such probabilities are
calculated assuming the oscillator at the inverse temperature $\beta$,
i.e. $p_j=\mathrm{Tr}_q[\varrho_q(\beta)\hat\Pi_j]$.  
\par
The quantum mechanical counterpart of the Fisher Information is defined as 
\begin{equation}
H(\beta)=\mathrm{Tr}[\varrho_q\hat{L}^2(\beta)]
\end{equation}
with $\hat L(\beta)$ the symmetric logarithmic 
derivative operator, satisfying the equation
\begin{equation}
\partial_\beta\varrho_q=[\hat L(\beta)\varrho_q+\varrho_q\hat L(\beta)]/2.
\end{equation}
The quantum Fisher Information (QFI) is an upper bound for $F(\beta)$ as it
embodies the optimization of the Fisher Information over any possible
measurement performed over the probing qubit states. The QFI 
is thus independent of the specific measurement strategy and
is an intrinsic feature of the family of probing states.
Eq.~(\ref{bound}) can then be rewritten as 
\begin{equation}
\delta(\beta)\ge\frac{1}{MH(\beta)},
\end{equation}
which extends the Cram\'er-Rao bound to the quantum domain and embodies
the ultimate limit to the precision of the estimate of $\beta$. A
measurement is optimal when the corresponding Fisher information 
$F(\beta)$ equals the quantum Fisher Information $H(\beta)$. 
Although various instances of optimal
measurement may be found, depending on the model at hand, the observable
embodied by the spectral measure of $\hat L(\beta)$ is certainly
optimal. Upon diagonalization of the probe state
$\varrho_q=\varrho_+\ket{\psi_+}\bra{\psi_+}_q+\varrho_-\ket{\psi_-}\bra{\psi_-}_q$,
the QFI can be computed explicitly as
\begin{align}
H(\beta)=&\sum_{k=\pm}\frac{(\partial_\beta\varrho_k)^2}{\varrho_k}
\notag \\
&+2\gamma\sum_{k\neq l=\pm}\left|
\sum_{j=0,1}(\partial_\beta\langle{j}|\psi_{k}\rangle)\langle\psi_l|j\rangle
\right|^2 
\end{align}
with $\gamma=(1-2\varrho_{+})^2$~\cite{Bru11}. In what follows, we
consider two exactly solvable models corresponding to interesting
physical situations and compute the QFI to assess
the ultimate precision in the estimation of temperature achievable by
{\em any} measurements performed on the states of the probing system. We
will compare such optimal performance to what is obtained through the
Fisher Information associated with population measurements of the probe,
i.e. for $\{\hat\Pi_j\}=\{\ket{0}\bra{0}_q,\ket{1}\bra{1}_q\}$. We show
that in some cases, population measurements are indeed optimal for the
estimation of temperature. 
\section{Jaynes-Cummings coupling beyond the rotating wave approximation}
\label{modello1}
\begin{figure}[b] 
\includegraphics[width=0.4\textwidth]{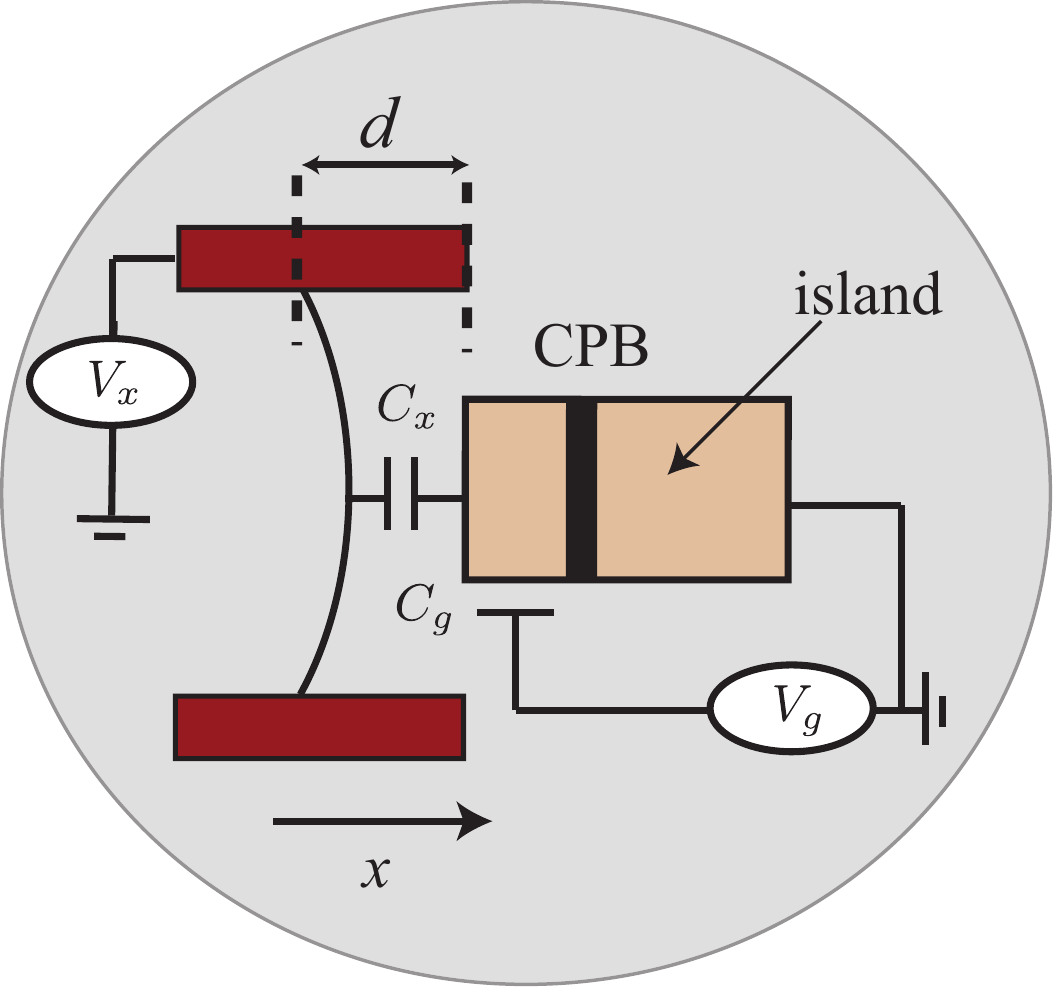} 
\caption{(Color online) An electrically driven nanomechanical oscillator (bias voltage
$V_x$) is coupled to a CPB through the capacitance $C_x$. The state of
the CPB is controlled by the gate voltage $V_g$ (coupled to the box
through the capacitance $C_g$) and the Josephson energy $E_J$. We work
at the charge degeneracy point.} \label{CPB} \end{figure}

The first model that we address corresponds to the choice ${\hat
A}_o\equiv\hat{X}_o=({\hat a+\hat a^{\dagger}})/{\sqrt{2}}$, i.e. the
in-phase quadrature operator of the harmonic oscillator, and $\hat
A_q=\hat\sigma_x$, which is the $x$-Pauli spin operator.
Correspondingly, the interaction reads 
$$\hat H_I= g\, \hat X_o\otimes \hat\sigma_x\,.$$ 
This model is encountered in a few different contexts. On
one hand, it describes the effective interaction Hamiltonian for the
electric-dipole coupling between a two-level atom and the field of a
cavity, thus embodying the celebrated Jaynes-Cummings
Hamiltonian~\cite{JC} beyond the so-called rotating wave approximation.
Moreover and rather less intuitively, the same model is achieved by
considering a nanomechanical oscillator (a {\it nano beam}) coupled
capacitively to a Cooper-pair box (CPB) operating at the so-called charge
degeneracy point~\cite{schon}, where the dynamics of the CPB can be
righteously be approximated to that of a two-level system encoded in the
space spanned by states $\ket{\pm}=(\ket{0}\pm\ket{1})/\sqrt{2}$.  Here
$\{\ket{0},\ket{1}\}$ are states with exactly 0 and $1$ excess Cooper
pairs in the large superconducting island shown in Fig.~\ref{CPB}. The
natural Hamiltonian of the system reads 
\begin{equation}
\hat H_1=\frac{(\hat Q-Q_g)^2}{2C_{t}}-E_J\cos\hat\phi+\Omega\hat a^\dag\hat a
\end{equation}
with $\hat Q$ and $\hat \phi$ the canonical charge and phase operator of
the CPB, $C_t$ the total capacitance of the  island, $Q_g=C_gV_g+C_xV_x$
the total gate charge, $E_J$ the Josephson energy and $\Omega$ the
frequency of the nanomechanical oscillator, as indicated
above~\cite{schon}.  
By defining $\hat\sigma_x=\ket{+}\bra{-}+\ket{-}\bra{+}$, expanding
$\hat H_1$  in series of the ratio $x/d$ between the actual position of
the mechanical oscillator and its equilibrium distance from the CPB (the
amplitude of the oscillations is assumed to be small enough that only
terms proportional to $x/d$ are retained in such expansion) and
adjusting the gate and driving voltages such that $Q_g\simeq0$, the
interaction Hamiltonian of the system can be cast into the form
\begin{equation} 
\label{ham1}
\hat H_1=\lambda(\hat a+\hat a^\dag)\otimes\hat
\sigma_x\equiv g\hat{X}\otimes\hat{\sigma}_x
\end{equation}
with $\lambda=g/\sqrt{2}$ an effective coupling rate whose 
form is inessential for our tasks.  
\par
The estimate of the temperature in this particular context is especially
relevant.  Indeed, the nano beam is in contact with a thermal phononic
background due to the substrate onto which it is
nano-fabricated~\cite{roukes}. The coupling with the superconducting
qubit addressed above holds the potential to prepare non-classical
states of the nano-beam. Indeed, the time-evolution operator
corresponding to Eq.~(\ref{ham1}) reads, in the qubit basis, as
$$\hat{U}(t)=\cos(gt\hat X)\openone_q-i\sin(gt\hat X)\hat\sigma_x\,.$$ 
Let
us assume that the nano-beam is initialized in a coherent state
$\ket{\alpha}~(\alpha\in\mathbb{R})$, while the qubit is prepared in
$\ket{+}_q$. The evolution will generate the qubit-oscillator state
\begin{equation}
\label{state}
\ket{\eta(\alpha)}_{qo}=\frac{1}{\sqrt 2}
(\ket{\alpha+igt}_o\ket{0}_q+\ket{\alpha-igt}_o\ket{1}_q).
\end{equation}
As $|\langle{\alpha-igt}|\alpha+igt\rangle|^2=e^{-4g^2t^2}$, for
$gt\simeq\pi$ the two coherent states $\ket{\alpha\pm igt}$ are
quasi-orthogonal and Eq.~(\ref{state}) is almost maximally entangled. By
projecting the qubit onto $\ket{+}_q$, we achieve the coherent-state
superposition ${\cal N}(\ket{\alpha+igt}_o+\ket{\alpha-igt}_o)$ (${\cal
N}$ is a normalization factor) which embodies, in the limit of
quasi-orthogonal coherent states mentioned above, a highly non-classical
state. However, a thermal-state preparation of the harmonic oscillator
will smear out such non-classicality, pushing the state towards the
statistical mixture 
\begin{equation}
\rho_{o,th}=\int\!\! d^2 \alpha\, G(\alpha,V)\,
|\eta(\alpha)\rangle_o{}_o\langle \eta(\alpha)|\,
\end{equation}
where $G(\alpha,V)$ is a Gaussian distribution of width
$V=2\nbar+1$~\cite{mauro05}. Determining the exact initial temperature
of the nano beam is thus key for the success of such conditional
strategies for the enforcement of non-classical features. Our approach
to the estimate of $\beta$ will follow the general strategy described
above, which we now describe for the specific model in Eq.~(\ref{ham1}).
\par
\begin{figure}[b] 
\includegraphics[width=0.21\textwidth]{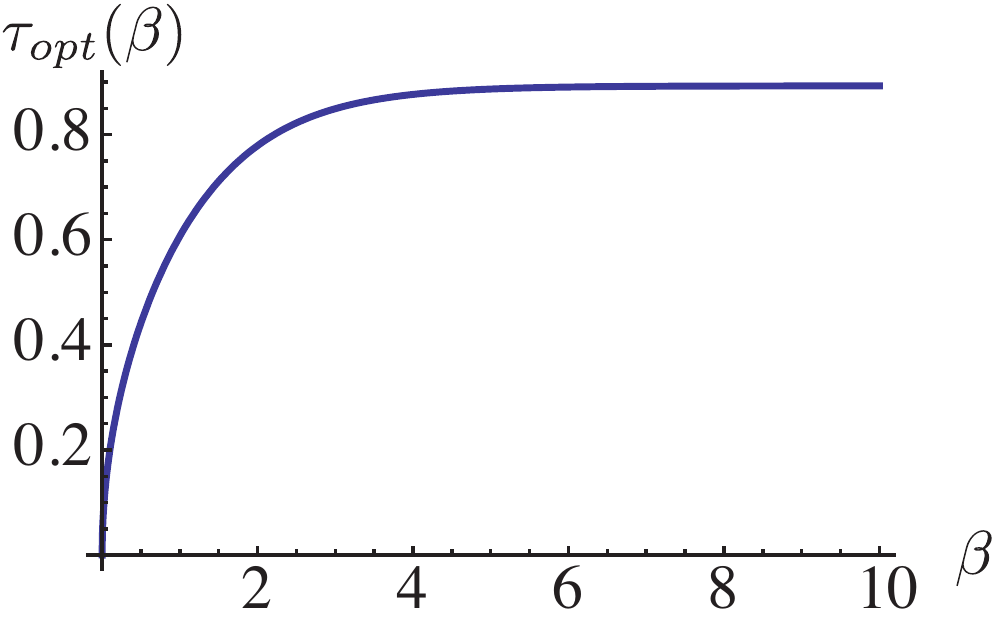}
\includegraphics[width=0.23\textwidth]{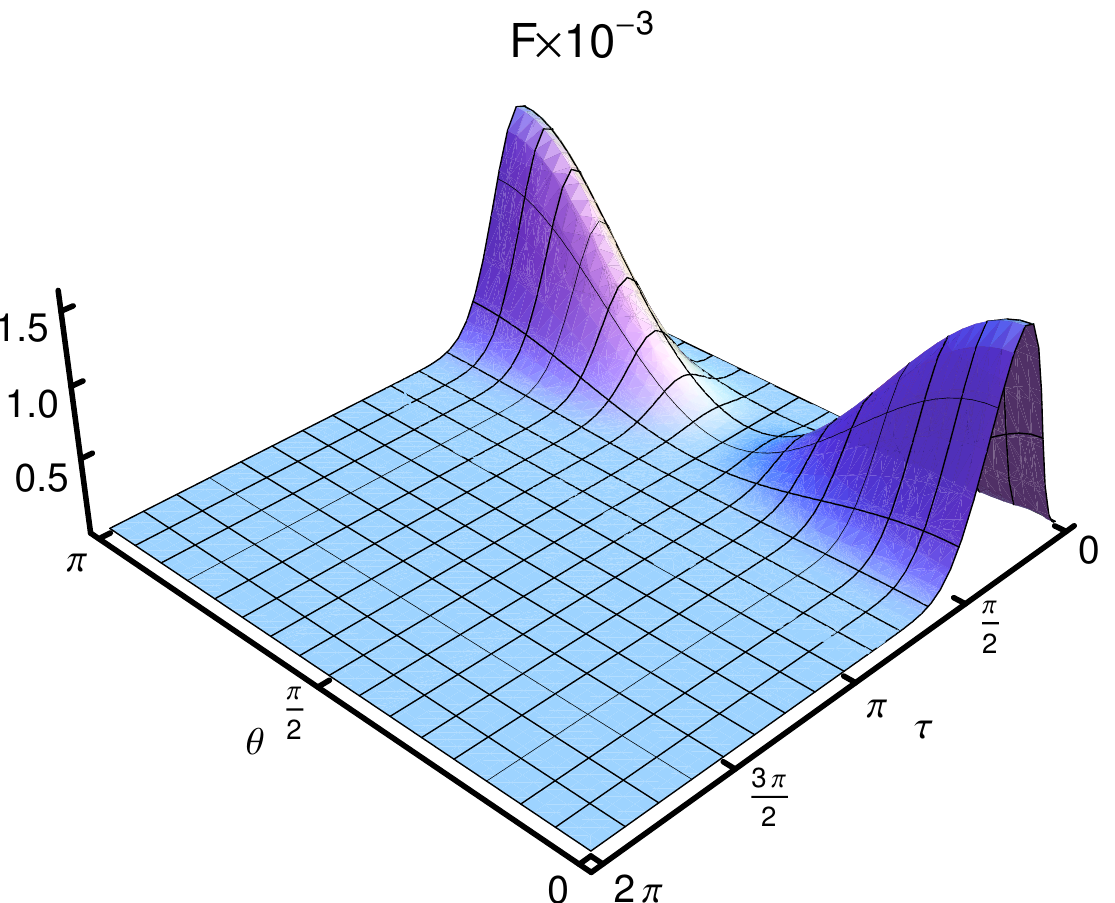} 
\caption{(Color online) Left: The functional form of $\tau_{opt}(\beta)$.
Right: $F(\beta=3)$ against $\theta$ and $\tau$. Clearly, the Fisher
Information is optimized at $\theta=0,\pi$ and quickly decays as $\tau$
grows. } \label{lambert}
\end{figure}

The elements of the state of the probing qubit after the interaction
with the oscillator and the trace over its degrees of freedom can be
calculated explicitly as 
\begin{equation}
\varrho_q(\beta){=}\frac12
\left[\begin{matrix}
1 + \cos \theta e^{-\zeta}
& {\sin \theta}( \cos \varphi -i \sin \varphi e^{-\zeta})\\
{\sin \theta}( \cos \varphi +i 
\sin \varphi e^{-\zeta})&1 - \cos \theta e^{-\zeta}
\end{matrix}\right]\label{varq}
\end{equation}
with $\zeta=\mathrm{\coth}\left( \frac{\beta}{2}\right)\tau^2$, where $\beta=\frac
{\Omega}{k_B T}$ and $\tau=gt$ are respectively the dimensionless inverse 
temperature and interaction time. The Fisher Information
associated with a measurement of the populations of $\varrho_q(\beta)$,
i.e. a measurement of the $z$-Pauli operator $\hat\sigma_z$, reads
\begin{equation*}
F(\beta)=\frac{\cos^2\theta \, \mathrm{\csch}^4\left( {\beta}/{2}\right)}
{e^{2 \zeta} - \cos^2\theta}
\frac{\tau^4}{4} \, .
\end{equation*}
which is a function of $\beta$, $\theta$, $\tau$.  Compared to the case
where the rotating wave approximation is invoked~\cite{Bru11}, i.e. for
a qubit-oscillator interaction of the form $g(\hat
a^\dag\hat\sigma_-+h.c.)$ with $\hat\sigma_\pm$ the ladder operators of
the qubit, the Fisher Information displays a symmetric behavior with
respect to $\theta$ and is no longer a periodic function of the time
$\tau$. The maximum is achieved by choosing $\theta=\{0,\pi\}$, i.e. by
preparing the qubit either in $\ket{0}$ or $\ket{1}$, while for
$\theta=\frac \pi 2$ the Fisher Information identically vanishes.
Finally, upon choosing one of the optimal qubit preparations, we found
that the maximum value of the Fisher Information only depends on
temperature. In fact, the value of the dimensionless time at which the
Fisher Information is maximized  is the following function of $\beta$
\begin{align}
\tau_{opt}(\beta)&=\sqrt{\left[1+\frac12W
\left(-\frac{2}{e^2}\right)\right]\tanh\left(\frac{\beta}{2}\right)}
\\ & \simeq  0.893 \times (1-e^{-\beta})\notag
\end{align}
with $W(y)$ the Lambert function of argument $y$~\cite{corless1996}.
The
behavior of $\tau_{opt}(\beta)$ is shown in the left panel of 
Fig.~\ref{lambert}, while in the right one we give examples of the form of the
$F(\beta)$ for large temperatures. Remarkably, in the relevant regime
$\beta\gg 1$ (low temperatures), the optimal interaction time becomes
almost independent on $\beta$. This means that no fine tuning of the
interaction time is needed and only a rough a priori information is
needed to implement the optimal measurement.  
\par
\begin{figure}[t!] 
\includegraphics[width=0.4\textwidth]{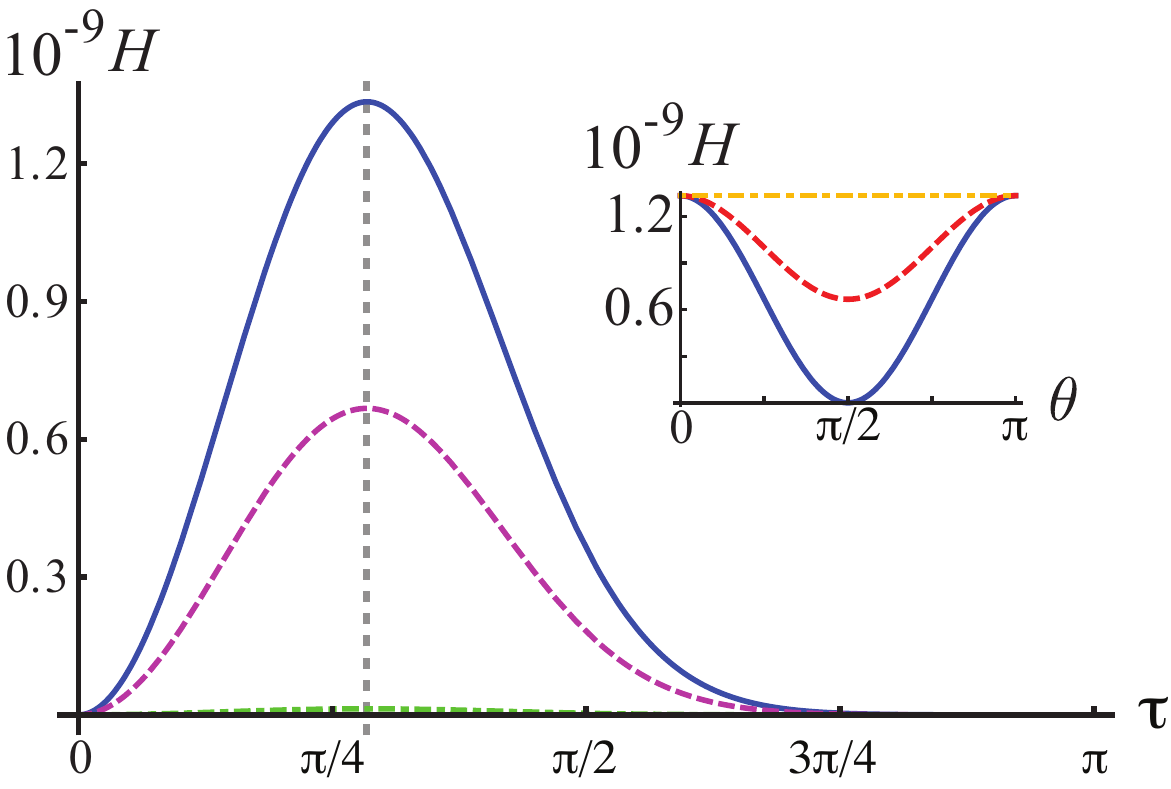} 
\caption{(Color online) Temporal evolution of the quantum Fisher Information for
$\beta=10$, $\varphi=0$ and $\theta=0$ (blue), $\theta=\frac{\pi}{4}$
(magenta), $\theta=\frac{\pi}{2}+0.1$ (green).  For $\theta=0$, the
quantum Fisher Information is maximum (independently on $\varphi$) and
equals the Fisher Information associated to population measurements,
while for $\theta=\frac{\pi}{2}$ $H(\beta)$ identically vanishes. Inset:
the quantum Fisher Information evaluated at the optimal time, plotted
against $\theta$ and for $\varphi=0$ (blue), $\varphi=\frac{\pi}{4}$
(red), $\varphi=\frac{\pi}{2}$ (orange). The last choice leads to the
same maximum value of the quantum Fisher Information and does not depend
on $\theta$.  \label{f:H_tau}}
\end{figure}

In order to evaluate the quantum Fisher Information, we have
diagonalized the state of the probe, as described in Sec.~\ref{modello}.
The explicit calculation, which produces expressions too involved to be
reported here, shows that $H(\beta)$ is maximized for two independent
sets of choices of the qubit-state parameters. One can either prepare
the qubit in one of the basis states $\ket{0}$ or $\ket{1}$,
independently of the angle $\varphi$, or choose $\varphi=\{  \pi/2,
3\pi/2 \}$, regardless of $\theta$ (cf. Fig. \ref{f:H_tau}). The values
of the QFI are the same in both cases, and the analytic expression of
$H(\beta)$ reduces to the one taken by $F(\beta)$ for the choice
$\theta=0$. This demonstrates that population measurements are optimal
for the whole range of temperatures. 
\par
This conclusion is further strengthened by the analysis of the spectral
measure of the symmetric logarithmic derivative $\hat L(\beta)$.  For
both the optimal probe-state preparations, $\hat{L}(\beta)$ is diagonal
and reads 
\begin{equation}
\label{spec1}
L(\beta)=-\frac{\tau^2}{4}\, \mathrm{\csch}^2\left(
\frac{\beta}{2}\right)\left[ \left(\mathrm{\coth}\zeta-1\right)
\hat\openone +\, \mathrm{\csch}\zeta \hat\sigma_z
\right].
\end{equation} 
with $\hat\sigma_z$ the $z$-Pauli operator. The explicit presence of
such operator in Eq.~(\ref{spec1}) demonstrates the optimality of
population measurements for the estimation of temperature in this model.
\par
We end the section by noticing that if the harmonic oscillator is
moved away from the equilibrium position, i.e. its initial state is
described by the displaced thermal state $D(\alpha) \varrho_o D^\dag
(\varrho)$, $D(\alpha) = \exp\{\alpha a^\dag -\bar\alpha a\}$ being the
displacement operator, then the probe qubit after the interaction is
given by  $$\varrho_q (\alpha,\beta)= 
e^{-i g \alpha \sigma_x \tau} \varrho_q (\beta) e^{i g \alpha \sigma_x
\tau}\,,$$ where $\varrho_q (\beta)$ is the probe state of Eq. (\ref{varq}).
The QFI is equal to the zero displacement case, whereas the Fisher
information is in general smaller than for zero displacement. 
\section{Far off-resonant spin-boson interaction}
\label{modello2}
We now address a second qubit-oscillator interaction model, specified by
taking $\hat{A}_o=\hat a^{\dagger}\hat a$ and $\hat A_q=\hat\sigma_x$.
The interaction Hamiltonian thus becomes 
\begin{equation}
\label{secondo}
\hat H_2=  g\, \hat a^{\dagger}\hat a\otimes
\hat \sigma_x.
\end{equation} 
This model describes for a two-level system interacting  far
off-resonantly with a bosonic mode. Let us consider a two-level system
(bosonic mode) with transition frequency $\omega$ ($\Omega$),
interacting through a Jaynes-Cummings model with strength $\lambda$. We
call $\Delta=\Omega-\omega$ the detuning between the two systems.  The
corresponding time-evolution operator can be written, in the basis
$\{\ket{1},\ket{0}\}$ of the two-level system, as~\cite{stenholm}
\begin{equation}
\hat U=
\left[
\begin{matrix}
\cos(\hat\Omega_{n+1}t)-\frac{i\Delta}2  \hat K_{1+n}
&
-i\lambda\hat a \hat K_n 
\\
i\lambda\hat a^\dag \hat K_n 
&
\cos(\hat\Omega_{n}t)+\frac{i\Delta}2 \hat K_n
\end{matrix}
\right]
\end{equation}
where 
$$
\hat K_n = 
\frac{\sin(\hat\Omega_{n}t)}{\hat\Omega_{n}}\,,
$$
$\hat\Omega_n=\sqrt{\frac14 \Delta^2+\lambda^2\hat a^\dag\hat a}$ 
is the effective Rabi frequency operator, and 
$\hat{a}$ ($\hat{a}^\dag$) are the field operators of the boson.
For
$\Delta^2/4\gg{\lambda^2\langle\hat a^\dag\hat a\rangle}$, we have 
\begin{align}
\hat U_{10}&=\hat U_{01}\simeq0 \notag \\ 
\hat U_{11}&\simeq e^{-i\hat\Omega_{n+1}t} \quad 
\hat U_{00}\simeq e^{i\hat\Omega_{n}t}\,, 
\end{align}
and by moving to a reference frame rotating at frequency $\Delta$, we
gain the effective picture~\cite{ogden06} 
\begin{equation}
\hat U\simeq e^{-i\frac{\lambda^2}{\Delta}\hat a\hat a^\dag t}
\ket{1}\bra{1}+e^{i\frac{\lambda^2}{\Delta}\hat a^\dag\hat a t}\ket{0}\bra{0}.
\end{equation}
We can now shift the energy of the two-level system so that $\ket{g}$
becomes the zero-energy state and $$\hat U=e^{-i\hat H'_2
t}=e^{-i\frac{\lambda^2}{\Delta}\hat a\hat a^\dag \ket{1}\bra{1}t}\,,$$ with
$\hat H'_2$ the interaction Hamiltonian between the boson and the
two-level system. By reminding that $\ket 1\bra1=2\hat\sigma_z+\openone$
and neglecting an inessential term depending only on the qubits' degrees
of freedom, we gather the non-trivial interaction term
$\frac{2\lambda^2}{\Delta}\hat a^\dag \hat a\otimes \hat\sigma_z$. This
is locally equivalent (via a Hadamard gate applied to the two-level
system) to the model $\hat H_2$ in Eq.~\eqref{secondo}. We now assume
this interaction model for the probe-oscillator dynamics and the
protocol for the estimate of the temperature.  
\par
The matrix elements of the probe state after evolution and the trace
over the oscillator are 
\begin{equation}
\begin{aligned}
&\varrho_{q,00}{=}\cos^2{\frac{\theta}{2}} +\Gamma[2\zeta \cos \theta\,
\mathrm{sinc}^2 \tau {-} \sin \theta \sin \varphi \sin(2\tau)],\\
& \varrho_{q,01}{=}\frac{\sin \theta\!\cos \varphi}{2} 
{+}i\Gamma
\{[e^\beta\!{-}\!\cos(2\tau)]{\sin \theta \sin \varphi}\!{-}\!\cos\theta\sin(2\tau)\}
\end{aligned}
\end{equation}
with $\varrho_{q,10}{=}\varrho_{q,01}^*$, $\varrho_{q,11}=1-\varrho_{q,00}$ and 
$$\Gamma=\frac{1{-}e^{-\beta}}{4\left[ \cos(2
\tau){-}\cosh\beta\right]}\,.$$
\par
\begin{figure}[h!] 
\includegraphics[width=0.45\columnwidth]{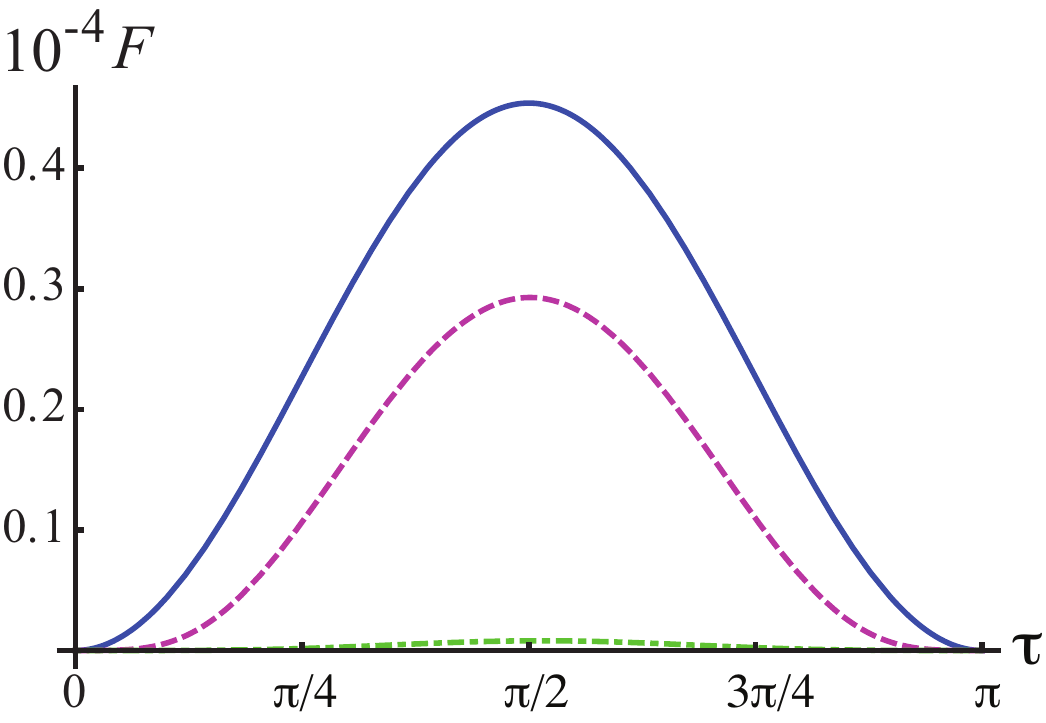} 
\includegraphics[width=0.47\columnwidth]{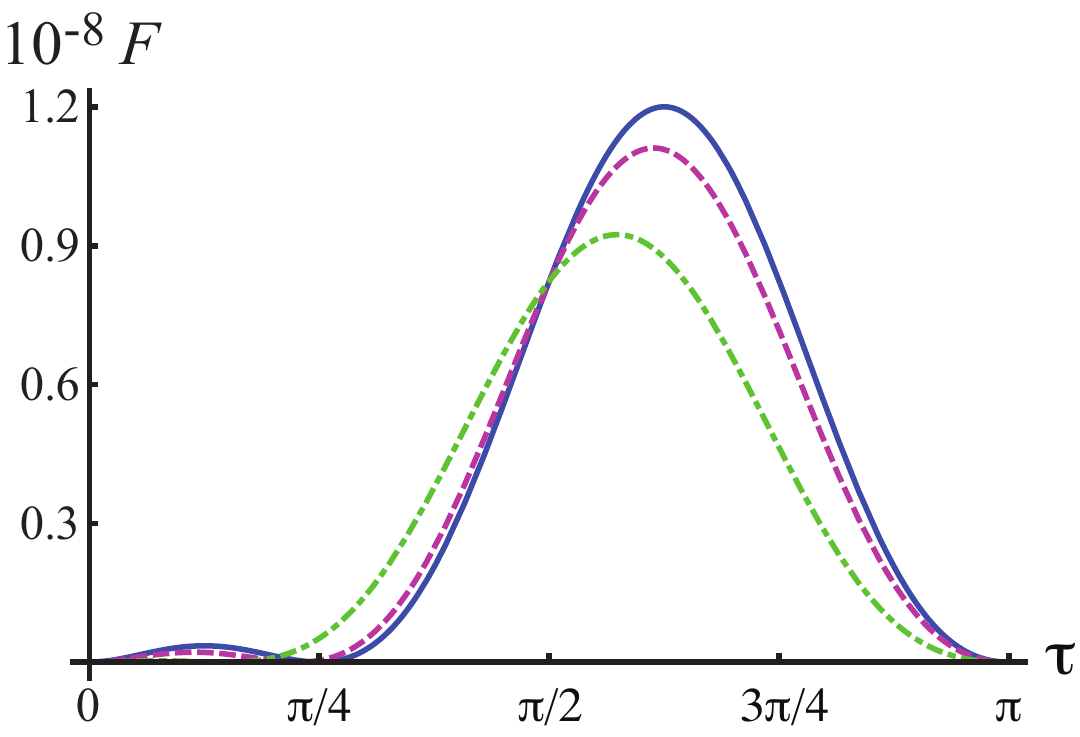} 
\caption{(Color online) Left: FI for $\beta=10$, $\varphi=\frac{\pi}{2}$  as a function of
the effective time $\tau$, for different $\theta$ values:
$\theta=0$ (blue), $\theta=0.01$ (magenta) and
$\theta=0.1$ (green). For $\theta=0$ the Fisher Information does 
not depend on $\varphi$. Right: Fisher Information for $\beta=10$
and $\theta=\frac{\pi}{4}$ as a function of time for different 
$\varphi$: $\varphi=\frac{\pi}{2}$ (blue), $\varphi=\frac{\pi}{3}$ 
(magenta) and $\varphi=\frac{\pi}{6}$ (green). Although the 
choice $\varphi=\frac{\pi}{2}$ maximizes the Fisher Information, it is 
evident that the relevant parameter in setting the 
qubit is $\theta$.
\label{f:fish}}
\end{figure}

The expression taken by the Fisher Information is, in this case, too
lengthy to be reported. Qualitatively, $F(\beta)$ depends on both
$\theta$ and $\varphi$ and, as in the Jaynes-Cummings model under the
rotating-wave approximation, is a periodic function of time $\tau$.
The probe state preparation that optimises the Fisher Information is again
$\theta=\{ 0,\pi\}$. For both such choices, $F(\beta)$ is independent of
$\varphi$. However, as soon as the qubit initialization deviates from
the optimal cases, $F(\beta)$ suddenly drops by several orders of
magnitude, as shown in the left panel of Fig.~\ref{f:fish}, displaying a weak
dependence on $\varphi$ [taking $\varphi=\frac \pi2$ gives the maximum
of $F(\beta)$, see the right panel of Fig.~\ref{f:fish}]. However, the values
attained by the Fisher Information in at such optimal values of
$\varphi$ are negligible with respect to those associated to
$\theta=\{0,\pi\}$ [there is a difference of four orders of magnitude
between the values in the left and the right panels of Fig.~\ref{f:fish}], 
which makes $\theta$ the only effective qubit parameter. 
\par
\begin{figure}[h!] 
\includegraphics[width=0.95\columnwidth]{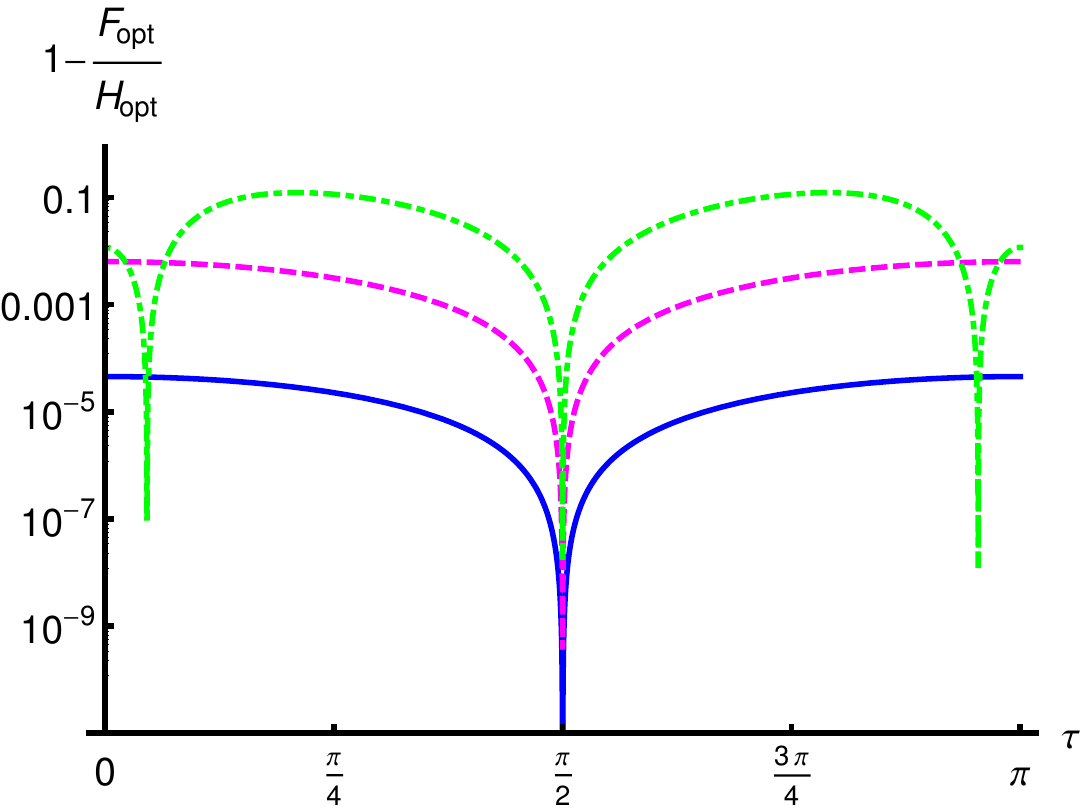} 
\caption{(Color online) Logarithmic plot of the quantity
$[H_{opt}(\beta){-}F_{opt}(\beta)]/H_{opt}(\beta)$ for $\beta=10$
(blue), $\beta=5$ (magenta) and $\beta=1$ (green).}  \label{f:comp}
\end{figure}

If we prepare the qubit in an eigenstate of $\hat\sigma_z$, the Fisher
Information associated to a population measurement reads
\begin{widetext}
\begin{equation}
F(\beta)_{opt}=\frac{2 e^{2 \beta}\sin^2\tau \left[ 1 +{\sinh}\beta - 
\cos(2\tau)e^{-\beta}\right]^2
{\tanh}\left(\frac{\beta}{2} \right)}{\left(e^{\beta}-1 \right)
\left[ 1 + (e^{\beta}-\cos(2\tau))\left(2e^{\beta} 
-1\right)-e^{\beta}\cos(2\tau)\right] 
\left[ \cos(2\tau)- {\cosh}\beta\right]^2}
\stackrel{\beta \gg 1}{\simeq} e^{-\beta} \sin^2 \tau
\end{equation}
\end{widetext}
On the other hand, by inspecting $H(\beta)$ we found, as before, a
symmetric behavior with respect to the qubit parameters: at a given
dimensionless time $\tau$, $H(\beta)$ is maximum either for $\theta=\{
0,\pi\}$ (regardless of $\varphi$), or $\varphi=\{ \frac \pi2, \frac32
\pi \}$ (regardless of $\theta$) and the values achieved by the QFI are
equal in both cases. Hence the effective contribution to the dynamics
comes from one octant of the Bloch sphere.  The analytic expression of
the optimal quantum Fisher Information is 
\begin{align}
H_{opt}(\beta)&=\frac{\sin^2\tau\left[ 2\cos(2\tau) - 2\,
\mathrm{\cosh}\beta - \, \mathrm{\sinh}^2\beta
\right]}{2\left[  \cos(2\tau) - \, \mathrm{\cosh} \beta
\right]^3}\notag \\ 
&\stackrel{\beta \gg 1}{\simeq} e^{-\beta} \sin^2 \tau\,.  
\end{align}
The crucial point here is that, for this model, the optimal Fisher
Information and its quantum mechanical counterpart are no longer the
same, i.e. population measurements are not the optimal one for the whole
set of parameters: at low temperatures, provided that we choose an
optimal qubit-state preparation, we retrieve the optimality of
$\hat\sigma_z$ measurements as $F_{opt}(\beta)=H_{opt}(\beta)$. When the
temperature is raised, on the other hand, small discrepancies appear
between the temporal behavior of these quantities, suggesting that $\hat
\sigma_z$ is not longer the best measurement strategy. This is shown in
Fig.~\ref{f:comp}, where we study the relative difference between
optimized Fisher and quantum Fisher Information. 
\par
Some insight comes from
the analysis of the symmetric logarithmic derivative, which reads
\begin{equation}
L(\beta)= L_0 \hat \openone - L_x \hat\sigma_z + L_z \hat \sigma_z
\end{equation} 
showing the presence of a contribution coming from a term proportional to
$\hat\sigma_x$, which is responsible for the differences between the two
estimators.
The expression of the coefficients reads as follows
\begin{align}
L_0=& \frac{\mathrm{\sinh}\beta}{2\left[ \cos(2\tau){-}
\mathrm{\cosh}\left(\beta \right)\right]} 
\stackrel{\beta \gg 1}{\simeq}
-\frac12 - e^{-\beta} \cos 2\tau \notag \\
L_x =& \frac{e^{\beta}\left(e^{\beta} - 1 \right) \left|
\sin(2\tau)\right|^2}{\left[1+e^{2\beta} {-} 2 e^{\beta}
\cos(2\tau)\right]^{3/2}}
\stackrel{\beta \gg 1}{\simeq}
e^{-\beta} \sin^2 2\tau \notag \\
L_z =&
\frac{e^{\beta}+ 1}{2[1+e^{2\beta} - 2 e^{\beta}
\cos(2\tau)]^{1/2}}
\stackrel{\beta \gg 1}{\simeq}
\frac12 + e^{-\beta} \cos^2 \tau\,,
\end{align} 
and shows explicitly that optimality of population measurement is
recovered for in the low temperature regime $\beta\gg 1$.
\section{Conclusions}
\label{conc}
We have addressed the thermometry of a (directly inaccessible) quantum
harmonic oscillator through its coupling to a quantum probe embodied by
a controllable qubit that can be subjected to any measurement. By
focusing our attention on two models of current physical relevance and
using the framework of the (quantum) estimation theory, we have
determined the preparation of the probe qubit, the measurement and the
value of the interaction time that optimize the estimate of the
oscillator's temperature. We found that population measurements performed over the
probing system are nearly optimal for an ample range of temperatures.
This quite important from the operational point of view, given the
handiness of implementing $\hat \sigma_z$ measurements in all of the
settings that have been explicitly addressed here. Our work thus aims at
proposing an experimentally viable pathway towards the quantum-limited
inference of the properties of inaccessible quantum systems,
demonstrating that the paradigm of the coupling with a (fully
controllable) low-dimensional quantum system is indeed effective. 
We are working
towards the extension of this framework to explicitly open-system
dynamics and the characterization of the environmental properties
affecting the dynamics of the harmonic oscillator. 
\acknowledgments
This work has been supported by MIUR (FIRB-RBFR10YQ3H-LiCHIS). MP thanks
the UK EPSRC for financial support through a Career Acceleration
Fellowship (EP/G04759/1) and a grant from the ``New Directions for EPSRC
Research Leaders" initiative. 

\end{document}